\documentclass[conference]{IEEEtran}
\ifCLASSINFOpdf
\else
\fi
\usepackage{graphicx}
\usepackage{tikz}
\usepackage{amsmath}
\usepackage{xcolor}
\usepackage{booktabs}
\usepackage{multirow}

\begin{document}
%
\title{TransTARec: Time-Adaptive Translating Embedding Model for Next POI Recommendation}

\author{\IEEEauthorblockN{Yiping SUN*}
\IEEEauthorblockA{School of Electronic Information\\ and Electrical Engineering\\ Shanghai JiaoTong University\\ Shanghai, China
}
}


%


\maketitle

\begin{abstract}
The rapid growth of location acquisition technologies makes Point-of-Interest(POI) recommendation possible due to redundant user check-in records. In this paper, we focus on \textit{next} POI recommendation in which next POI is based on previous POI. 
We observe that time plays an important role in next POI recommendation but is neglected in the recent proposed translating embedding methods. To tackle this shortage, we propose a time-adaptive translating embedding model (TransTARec) for next POI recommendation that naturally incorporates temporal influence, sequential dynamics, and user preference within a single component. Methodologically, we treat a (previous timestamp, user, next timestamp) triplet as a union translation vector and develop a neural-based fusion operation to fuse user preference and temporal influence. The superiority of TransTARec, which is confirmed by extensive experiments on real-world datasets, comes from not only the introduction of temporal influence but also the direct unification with user preference and sequential dynamics. 
\end{abstract}


%
\IEEEpeerreviewmaketitle

\section{Introduction}
With the proliferation of location-acquisition technologies, users can record their check-in data and post them on the location-based social networks(LBSN). The vast amounts of check-in records enable personalized POI recommendation task, which has attracted a lot of research\cite{DBLP:conf/kdd/YangBZY017, DBLP:conf/cikm/MaZWL18}.  Compared to POI recommendation, \textit{next} POI recommendation is a particular case that predicts the next POI of the user based on the previous POI. The task allows people to understand their mobility behaviors better and help people explore the new appealing places.
In this paper, we would like to focus on \textit{next} POI recommendation.
In formal, we can describe relation classification as follows: Given a natural sentence $x_1,x_2,...,x_n$ with two entities $e_1,e_2$, the system needs to classify the semantic relation $y$ between two entities.

\begin{figure}[t]
    \vspace*{0.2in}
    \setlength{\belowcaptionskip}{-0.6cm}
    \centering
    \includegraphics[width=0.46\textwidth]{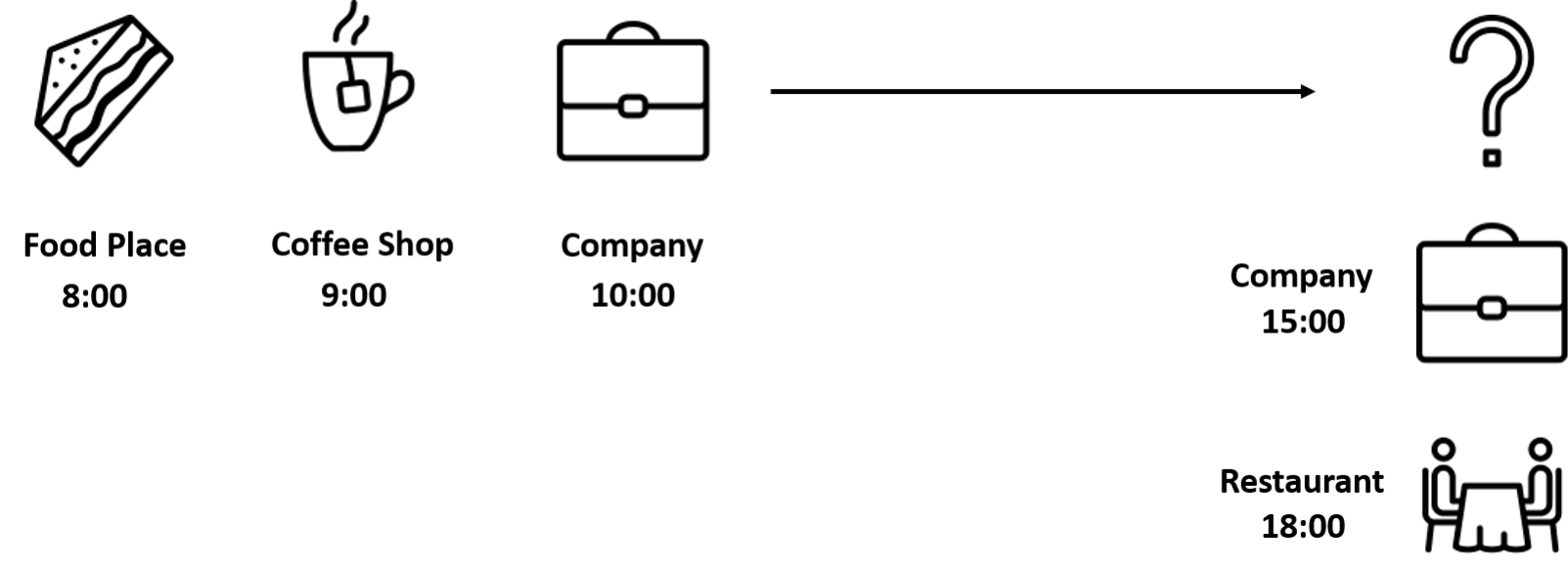}
    \caption{Toy example of user check-in sequences. }
    \label{fig:problem example}
\end{figure}

Time is of importance in next POI recommendation task. As illustrated by the toy example in Fig.[\ref{fig:problem example}]: the user may still work at 15:00 but go to a restaurant at 18:00. It is natural that a person maybe has different POI intentness at different time. 
Further, unlike other recommendation tasks that have continuous user activity records (i.e., user click profile), next POI recommendation task gets incomplete user records owing to the privacy protection. 
The challenges for considering temporal influence in next POI recommendation lie in two aspects: (1) How to jointly model temporal influence, sequential dynamics, and user preference? (2) How to deal with large scale and data sparsity problem when introducing temporal influence? 

For the first challenge, it was partially solved by TransRec\cite{ijcai2018-734}. They made such a translating operation to unify sequential dynamics and user preference: the embedding of the previous POI $p_i$, plus the translation vector of $u$ determines (approximately) the embedding of the next POI $p_j$, i.e., $\mathbf{v}_{p_i}+\mathbf{v}_u\approx\mathbf{v}_{p_j}$. However, the temporal influence is neglected. Inspired by this work, we smoothly add temporal influence to translation vector so that we can consolidate temporal influence, sequential dynamics, and user preference within a single translation operation. For the second challenge, when introducing temporal influence to translation vector, the number of (previous timestamp, user, next timestamp) triplets becomes tremendous (i.e $4.4$ billion in \textit{Foursquare} dataset). Traditional method (i.e. collaborative filtering) can only deal with pairs (triplets in our task). It is also difficult to update parameters jointly with the whole model. In our proposed method, a neural-based fusion operation is employed to solve large scale and data sparsity problem by mapping features into a latent space.

In this paper, we propose a time-adaptive translating embedding model \textit{TransTARec} (\textbf{\underline{T}}ime-\textbf{\underline{A}}daptive \textbf{\underline{Trans}}lation Embedding Model for Next POI \textbf{\underline{Rec}}ommendation).  Technically, we first treat a (previous timestamp $t_i$, user $u$, next timestamp $t_j$) triplet as a union translation vector $\mathbf{v}_{u,t}$ and develop a neural-based fusion approach to fuse user preference and temporal influence.  Then we make such a translation operation: the embedding of the previous POI $p_i$, plus the time-adaptive translation vector $\mathbf{v}_{u,t}$ should be approximately equal to the embedding of next POI $p_j$, i.e., $\mathbf{v}_{p_i} + \mathbf{v}_{u,t} \approx \mathbf{v}_{p_j}$.
Finally, we project the embedding of previous and next POI to the hyperplane of translation vector to deal with space inconsistency and complex relations issues\cite{wang2014knowledge} brought by translating embedding model. The main contributions of this paper can be
summarized as follows:
\begin{itemize}
    \item A new time-adaptive translating embedding model, named TransTARec, is proposed for next POI recommendation task. TransTARec extends the previous translating embedding model by exploiting temporal influence. 
    \item A neural-based fusion operation is proposed to fuse user preference and temporal influence. It helps to embed temporal influence in the translation vector and effectively tackles the large scale and data sparsity problem of translation vector.
    \item We have done extensive experiments on next POI recommendation over \textit{Foursquare} and \textit{Mobile} Dataset. The result shows that TransTARec outperforms previous embedding learning techniques from $8.41\%$ to $14.63\%$ on the precision of $Top@k$.
\end{itemize}

\section{Related Work}
In the literature, various methods have been proposed for next POI recommendation in recent years \cite{rendle2010factorizing,feng2015personalized}. Also, there are studies taking temporal influence into consideration for the purpose of improving the recommendation performance or other important fields like knowledge graph\cite{xiong2024teilp,xiong2024tilp}. For example, ST-DME\cite{ding2018spatial} proposed to incorporate geographical association and temporal periodic pattern via two pairwise distance metric embeddings. However, they model the two features separately, thus losing the generalization ability of implicit metricity. CARNN\cite{liu2016context} proposed to combine temporal influence and sequential dynamics via the recurrent neural network. Unfortunately, the recurrent neural network has difficulty in building the long-term dependency in next POI recommendation task because the length of user check-in records is always very long. Besides, in traditional POI recommendation task, STA\cite{Qian:2019:SRL:3306215.3295499} is similar to our method. They got the idea that the appearance of a user in some place at some time may indicate the POI. STA treated a (timestamp $t$, location $l$) pair as translation vector to operate on users $u$ and POIs $p$, i.e, $\mathbf{v}_u+\mathbf{v}_{t,l}\approx \mathbf{v}_p$. However, the lack of sequential behaviors makes it unsuitable to next POI recommendation task. On the other hand, TransTARec enjoys the implicit metricity via the unification of temporal influence, sequential dynamics and user preference by a single component. TransTARec is also easy to extend to large scale and long sequence scenario, thus making TransTARec more proper than previous methods. 

Artificial Intelligence has garnered considerable attention \cite{li2024feature,gao2023autonomous} in recent years. The techniques employed in this paper, such as neural-based methods \cite{yang2019neural,tang2024accelerating}, representation learning \cite{chen2023recontab}, information augmentation \cite{zhang2023optimizing}, and latent space embedding \cite{yao2023ndc}, are widely utilized across various domains. By integrating these techniques into our approach, we have achieved superior results compared to other methods. 
\section{Methodology}
\subsection{Problem Formulation}
TransTARec explicitly claims the temporal influence on next POI recommendation system. For each user $u\in \mathcal{U}$, we have a sequence of records $S^u = (S^u_1,S^u_2,...,S^u_{|S^u|})$ where each record refers to a (timestamp, POI) pair, i.e, $S^u_i= (t^u_i,p^u_i)$. Given the previous visited (timestamp $t_i$, POI $p_i$) pair and next timestamp $t_j$, our goal is to predict the next POI $p_j$ to visit by each user and generate recommendation lists as well. 

\subsection{The Proposed Model}
\begin{figure}[t]
    \centering
    \includegraphics[width=0.47\textwidth]{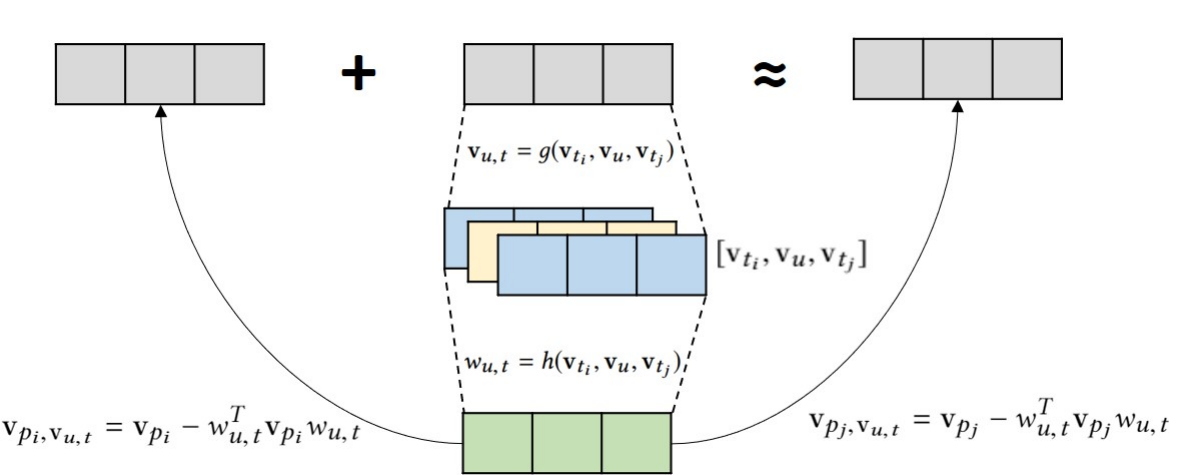}
    \caption{TransTARec Model: the embedding of previous POI $p_i$ is translated to the embedding of next POI $p_j$ via time-adaptive translation vector $\mathbf{v}_{u,t}$.}
    \label{fig:model}
\end{figure}
We aim to develop a time-adaptive translating embedding model that (1) naturally captures time-adaptive personalized sequential behavior and (2) easily deals with the large-scale and data sparsity problems of translation vectors. To achieve this goal, we need to obtain vectors in a high-dimensional latent space that accurately represent the semantic expression of (timestamp, POI) pairs. These vectors will enable us to utilize the translating embedding model for semantic transformation. Therefore, our main task is to train a model suitable for converting features into vectors and to construct a corresponding high-dimensional plane. The novelty of this proposed algorithm stems from two aspects. Firstly, while previous methods like TransRec only consider the user and item parts, we have designed a model with neural fusion that can incorporate important features like time into the entire translating model. Secondly, unlike other pure neural network methods, the translating embedding model can be easily scaled and offers high performance, as the main calculations primarily involve several vector operations. Compared to other methods that require numerous features, our approach can eliminate the matrix operations of large models. Moreover, we can simply add more features without increasing the complexity of the model. 

The detailed model explanation is in Fig.[\ref{fig:model}]. When we have a pair of (timestamp, POI) for a user, we transform the POI into a vector in a high-dimensional space. Then, we fuse it with the current timestamp, user information, and the next timestamp to obtain a fused high-dimensional vector. The following is how we aim to obtain the high-dimensional vector for the next POI for that user. Methodically, we first follow previous translating embedding methods TransRec to build the two-field translating embedding model in the translation space $\Phi$: the embedding of the previous POI $p_i$, plus the personalized translation vector $\mathbf{v}_u$ approximately determines the embedding of the next POI $p_j$, i.e, 
\begin{equation}
    \mathbf{v}_{p_i} + \mathbf{v}_{u} \approx \mathbf{v}_{p_j}
\end{equation}
In previous methods, $\mathbf{v}_{u}$ is a single preference vector for each user without considering other dimensions. In this paper, TransTARec extends it to the time-adaptive translation vector $\mathbf{v}_{u,t}$, which integrates the influence of previous timestamp $t_i$ and next timestamp $t_j$, i.e., 
\begin{equation}
   \mathbf{v}_{p_i} + \mathbf{v}_{u,t} \approx \mathbf{v}_{p_j}
\end{equation}
\begin{equation}
   \mathbf{v}_{u,t}=g(\mathbf{v}_{t_i},\mathbf{v}_{u},\mathbf{v}_{t_j}) = w_g[\mathbf{v}_{t_i},\mathbf{v}_{u},\mathbf{v}_{t_j}]+b_g
\end{equation}
where $[\cdot]$ is the concatenation operation and $t$ is the pair $(t_i,t_j)$ for simplicity. We treat a (previous timestamp, user, next timestamp) triplet as a union transition vector and develop a neural-based fusion operation (fully-connected layer) to generate translation vector adaptively. Note that our translation vector can deal with large scale and data sparsity problem of translating vector $\mathbf{v}_{u,t}$ due to the introduction of neural fusion. Next, we transform the absolute timestamp into (month $tm$, weekday $tw$ and hour $th$), then the time embedding is accumulated by the embedding of month, weekday and hour:
\begin{equation}
    \mathbf{v}_{t_i}=\mathbf{v}_{tm_i}+\mathbf{v}_{tw_i}+\mathbf{v}_{th_i}
\end{equation}
Inspired by TransH\cite{wang2014knowledge} in knowledge representation learning, we further project the vector of POI to the hyperplane of translation vector, which can drag vectors to the same hyperplane and deal with complex $1$-$N$, $N$-$1$, $N$-$N$ relations. We expect the projection $\mathbf{v}_{p_i,\mathbf{v}_{u,t}}$ and $\mathbf{v}_{p_j,\mathbf{v}_{u,t}}$ can be connected by a translation vector $\mathbf{v}_{u,t}$ on the hyperplane with low error if $\{(p_i,t_i),u,(p_j,t_j)\}$ is a golden triplet.
\begin{equation}
    \mathbf{v}_{p_i,\mathbf{v}_{u,t}} + \mathbf{v}_{u,t} \approx \mathbf{v}_{p_j,\mathbf{v}_{u,t}}
\end{equation}
By restricting norm vector $\|w_{u,t}\|_2^2=1$, it is easy to get the projection on the hyperplane of $\mathbf{v}_{u,t}$:
\begin{equation}
    \mathbf{v}_{p_i,\mathbf{v}_{u,t}}  = \mathbf{v}_{p_i} - w_{u,t}^T\mathbf{v}_{p_i}w_{u,t}
\end{equation}
\begin{equation}
    \mathbf{v}_{p_j,\mathbf{v}_{u,t}}  = \mathbf{v}_{p_j} - w_{u,t}^T\mathbf{v}_{p_j}w_{u,t}
\end{equation}
The norm vector is also calculated by the same neural-based fusion operation but with different weights and biases.
\begin{equation}
    w_{u,t} = h(\mathbf{v}_{t_i},\mathbf{v}_{u},\mathbf{v}_{t_j}) = w_h[\mathbf{v}_{t_i},\mathbf{v}_{u},\mathbf{v}_{t_j}]+b_h
\end{equation}
Finally, we define a score function to measure the plausibility whether the triplet is correct or not. The score is expected to be low for a golden triplet and high for an incorrect triplet. 
\begin{equation}
    f_r(p_i,p_j)=\|\mathbf{v}_{p_i,\mathbf{v}_{u,t}}+\mathbf{v}_{u,t}-\mathbf{v}_{p_j,\mathbf{v}_{u,t}}\|_2^2
\end{equation}

\subsubsection{Optimization and Learning the parameters}
Given a user and his records, the ultimate goal of the task is to rank the ground-truth POI $p_j$ at time $t_j$ higher than all other POIs($p_{j'}, j'\in \mathcal{I} \backslash j$). We follow the previous methods to optimize the pairwise ranking between $p_j$ and $p_{j'}$ using margin-based ranking loss. 
The following constraints are considered when we minimize the loss.
\begin{align}
    \forall u\in\mathcal{U},t\in\mathcal{T} &, \|w_{u,t}\|_2^2=1 \ \ \ // unit \  normal\  vector \\
    \forall u\in\mathcal{U},t\in\mathcal{T} &, |w_{u,t}^T\mathbf{v}_{u,t}|^2/\|\mathbf{v}_{u,t}\|_2^2\leq\epsilon^2 //orthogonal
\end{align}
The second constraint makes sure that $\mathbf{v}_{u,t}$ is in the hyperplane. To this end, we optimize the following loss function:
\begin{equation}
\begin{split}
    \mathcal{L} = & \sum_{u\in U}\sum_{(t_i,p_i)\in S^u}\sum_{(t_j,p_j)\in S^u}\sum_{(t_j,p_{j'})\notin S^u}[f_r(p_i,p_j)+\gamma\\&- f_r(p_i,p_{j'})]_+ + C(\sum_{u\in \mathcal{U},t\in\mathcal{T}}[\frac{|w_{u,t}^T\mathbf{v}_{u,t}|^2}{\|\mathbf{v}_{u,t}\|_2^2}-\epsilon^2]_+)
\end{split}
\end{equation}
where $[x]_+=max(0,x)$  and $C$ is the hyper-parameter weighting the importance of soft constraint. User embedding ($\mathbf{v}_u, \forall u\in\mathcal{U}$), POI embedding ($\mathbf{v}_p, \forall p\in\mathcal{P}$) and Time embedding ($\mathbf{v}_{tm}, \mathbf{v}_{tw}, \mathbf{v}_{th},$ $\forall tm,tw,th\in\mathcal{TM,TW,TH}$) are initialized randomly. To make the training phase efficient, negative sampling is applied when sampling the negative next POI $p_{j'}$. We realize the $\|w_{u,t}\|_2^2=1$ constraint by projecting each $w_{u,t}$ to unit $l_2$-ball before calculating the projection of POI embedding. 


\subsubsection{Ranking POIs for Recommendation}
Once all embedding vectors are learnt, we generate the next POI recommendation as follows. Given the previous POI $p_i$, previous timestamp $t_i$ and next timestamp $t_j$ of user $u$, we use the function to rank a candidate $p\in\mathcal{P}$.
\begin{equation}
\begin{split}
    R(u,t_i,p_i,t_j) & \propto (\mathbf{v}_{p_i,\mathbf{v}_{u,t}}+\mathbf{v}_{u,t})^T\cdot\mathbf{v}_{p,\mathbf{v}_{u,t}}\\
    & \propto (\mathbf{v}_{p_i,\mathbf{v}_{u,t}}+g(\mathbf{v}_{t_i},\mathbf{v}_{u},\mathbf{v}_{t_j}))^T\cdot\mathbf{v}_{p,\mathbf{v}_{u,t}}
\end{split}
\end{equation}
Note that when $j=i+1$ then the task is next POI recommendation but when $j=i+1, i+2,..., n$ the task can be switched to time-specific POI recommendation.

\section{Experiment}
\subsection{Experimental Settings}
\begin{figure}[htbp]
    \centering
    \includegraphics[width=0.3\textwidth]{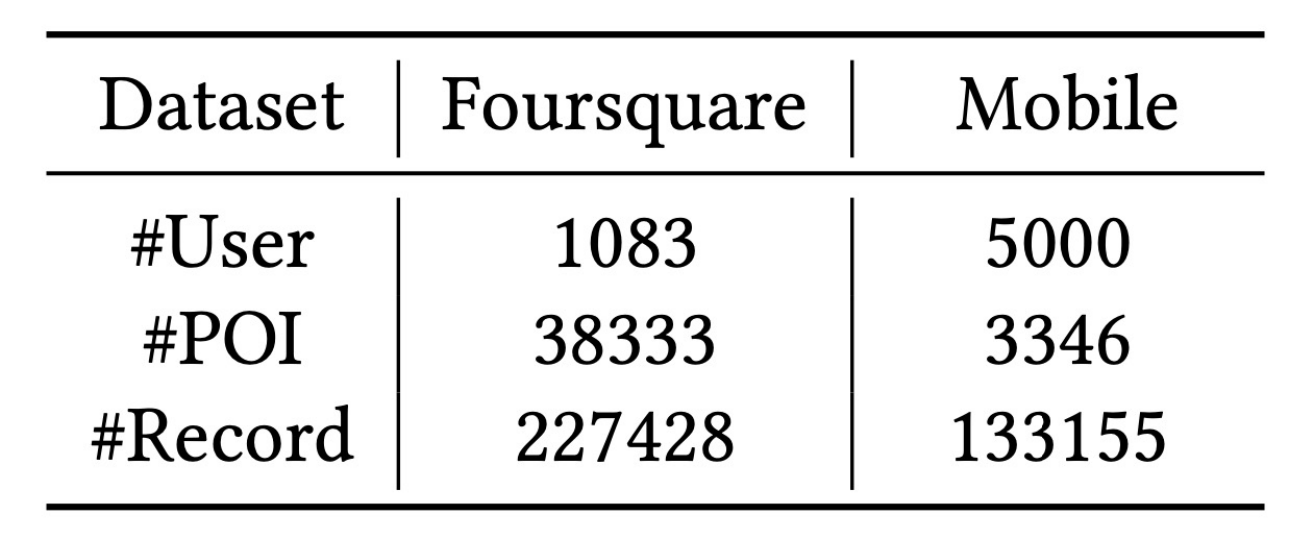}
    \caption{Dataset Analysis}
    \label{tab:commands}
\end{figure}
\begin{figure*}[htbp]
    \centering
    \includegraphics[width=0.9\textwidth]{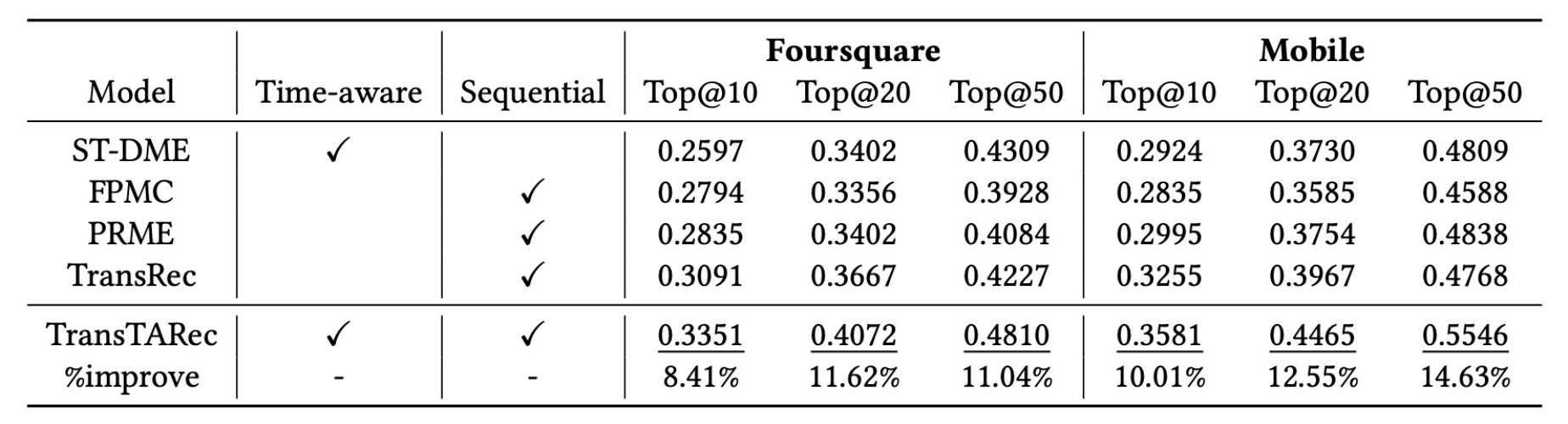}
    \caption{Result of Next POI Recommendation on \textit{Foursquare} and \textit{Mobile} Dataset}
    \label{tab:result}
\end{figure*}
Our experiments are conducted on two datasets. The \textit{Foursquare} dataset\cite{yang2015modeling} contains check-in data in NYC from 12 April 2012 to 16 February 2013. The \textit{Mobile} dataset contains user mobile roaming records from October 2018 to December 2018. The statistic analysis of two datasets is listed on Figure.[\ref{tab:commands}]. \textit{Foursquare} has more periodic data while \textit{Mobile} owns more continuous sequential data. We use the first $80\%$ of data as training data and the next $20\%$ of data as testing data. The hyperparameter is set as follows: dimension of embedding $d=100$, margin $\gamma=1.0$, soft constraint weight $C=1.0$ and auxiliary threshold $\epsilon=0.001$. 
We report the performance of each method on the test set in terms of the following ranking metrics: \begin{equation}
    Top@k=\frac{\sum_{u\in\mathcal{U}}\sum_{i\in u_{test}}\mathbf{1}(R_{u,i,g_{u,i}}\leq k)}{\sum_{u\in\mathcal{U}}\sum_{i\in u_{test}}\mathbf{1}}
\end{equation}
where $g_{u,i}$ is the ground truth of $i$-th test sample of user $u$.

\subsection{Performance Comparison and Analysis}
First, we conduct experiments on comparing proposed TransTARec model with state-of-the-art embedding learning methods: 
\begin{itemize}
     \item ST-DME\cite{ding2018spatial}: Incorporate geographical association and temporal periodic pattern via distance metric embedding. 
    \item FPMC\cite{rendle2010factorizing}: Use a predictor that combines Matrix Factorization and factorized Markov Chains so that personalized Markov behavior can be captured. 
    \item PRME\cite{feng2015personalized}: Propose to improve FPMC by learning two metric spaces: one for measuring user-item affinity and another for sequential continuity.
    \item TransRec\cite{ijcai2018-734}: Unify user preferences and sequential dynamics with translations.
\end{itemize}

Several observations are made from the result in Figure.[\ref{tab:result}]: (1) It is obvious that TransTARec outperforms previous matrix factorization, Markov Chains, metric embedding and translating embedding methods. The improvement comes from not only the introduction of temporal influence but also the direct unification with user preference and sequential dynamics. It shows that TransTARec is effective for next POI recommendation problem.  (2) Time-aware method ST-DME has a competitive $Top@50$ result, especially in \textit{Foursquare}. It proves that temporal influence helps to capture useful information. (3)PRME performs better over $Top@50$ on \textit{Mobile} dataset, proving that sequential dynamics plays an important role in the dataset that contains continuous sequential records. (4)TransRec is the state-of-the-art method that has the best $Top@10,Top@20$ results but sometimes lose on $Top@50$, showing the benefit of translating embedding methods on precision but reflecting the shortage of lacking additional information. 

\begin{figure}
    \centering
    \includegraphics[width=0.4\textwidth]{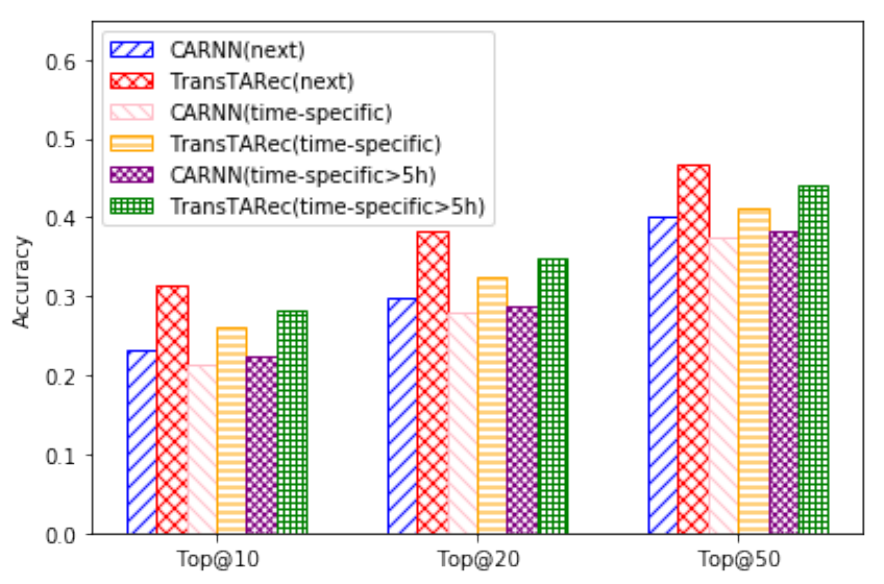}
    \caption{Time-specific POI Recommendation on Mobile$_{\leq100}$ Dataset}
    \label{fig:anytime}
    \vspace{-0.5cm}
\end{figure}

Then, we compare TransRec with recently proposed context-aware sequential recommendation model (CARNN)\cite{liu2016context} in deep learning. CARNN has the ability to combine temporal influence and sequential behavior together, which is competitive to our methods. They split the timestamp into time slot and use recurrent neural network to derive the next POI: $Pr(p_j|u,t,p)=f(u,t_1,p_1,...,t_i,p_i,t_j)$. We conduct three tasks (next POI recommendation, time-specific POI recommendation and time-specific (>5H) POI recommendation) by setting $j=i+1$; $j=i+1,i+2,...,n$; $j=i+p, \forall_p \ t_{i+p}-t_i\geq 5 \ hours$ respectively. Since CARNN is difficult to deal with long sequences, we regenerate a dataset that the length of user records is fewer than 100 from \textit{Mobile} Dataset. The result is shown in Fig.[\ref{fig:anytime}]. Firstly, it is clear that TransTARec can perform better than CARNN in all three tasks because of the generalization capability of embedding based methods. Secondly, CARNN still has competitive results due to its strong feature extraction ability. Thirdly, when the interval to the current time is longer than a threshold(5 hours in our experiment), the improvement expands to a larger margin, showing that TransTARec has utilized temporal influence better than CARNN.

Finally, we give a detailed case study of how recommendations vary according to the change of time. We randomly choose a user from \textit{Foursquare} Dataset, who seems to be a student. We select four timestamps as query and analyze the change of recommendation rank from such three places: Food Place, Home, and University. It shows that: (1) When at noon, Food Place will rank at the top and when at night, Home will rank at the top. (2)When the weekend comes, the university will not be the prior recommendation. (3)When the month has changed from May to June but when the absolute time in a day is not changed, the recommendation rank doesn't change much. The above three observations prove the effectiveness of the introduction of temporal influence by TransTARec.
\begin{figure}[htbp]
    \centering
    \includegraphics[width=0.47\textwidth]{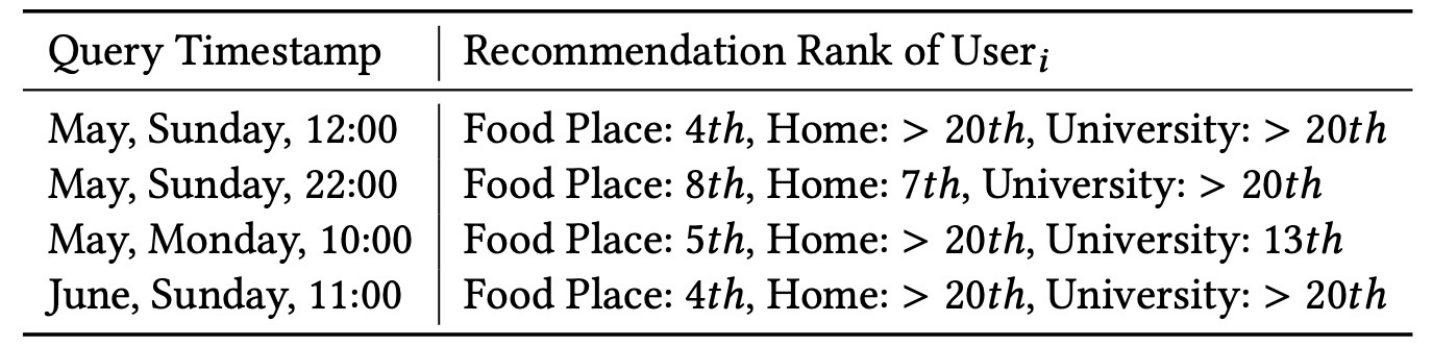}
    \label{tab:analysis}
    \caption{Case Study: Temporal Analysis of TransTARec}
\end{figure}
\section{Conclusion}
In this paper, we figure out the problem that temporal influence is ignored in translating embedding method and propose a time-adaptive translating embedding model for next POI recommendation (TransTARec). The unification of user preference, sequential behavior, and temporal influence brings TransTARec a better recommendation performance. The experiments show that TransTARec can outperform previous embedding based methods in a large margin and exploit temporal influence better than deep learning methods. Future work may utilize much more information (i.e., knowledge base) besides temporal influence.

\bibliographystyle{IEEEtran}
%



\end{document}